# AUXETIC-LIKE METAMATERIALS AS NOVEL EARTHQUAKE PROTECTIONS

**Bogdan Ungureanu[1,2*], Younes Achaoui[2*], Stefan Enoch[2], Stéphane Brûlé[3], Sébastien Guenneau[2]**

[1] *Faculty of Civil Engineering and Building Services Technical University "Gheorghe Asachi" of Iasi, 43, Dimitrie Mangeron Blvd., Iasi 700050, Romania,*
[2] *Aix-Marseille Université, CNRS, Centrale Marseille, Institut Fresnel UMR7249, 13013 Marseille, France,*
[3] *Dynamic Soil Laboratory, Ménard, 91620 Nozay, France.*

Email: bogdan.ungureanu@fresnel.fr ; sebastien.guenneau@fresnel.fr

*Equal contributing authors

**Abstract.** We propose that wave propagation through a class of mechanical metamaterials opens unprecedented avenues in seismic wave protection based on spectral properties of auxetic-like metamaterials. The elastic parameters of these metamaterials like the bulk and shear moduli, the mass density, and even the Poisson ratio, can exhibit negative values in elastic stop bands. We show here that the propagation of seismic waves with frequencies ranging from 1Hz to 40Hz can be influenced by a decameter scale version of auxetic-like metamaterials buried in the soil, with the combined effects of impedance mismatch, local resonances and Bragg stop bands. More precisely, we numerically examine and illustrate the markedly different behaviors between the propagation of seismic waves through a homogeneous isotropic elastic medium (concrete) and an auxetic-like metamaterial plate consisting of $4^3$ cells (40mx40mx40m), utilized here as a foundation of a building one would like to protect from seismic site effects. This novel class of seismic metamaterials opens band gaps at frequencies compatible with seismic waves when they are designed appropriately, what makes them interesting candidates for seismic isolation structures.

## 1. Introduction

Auxetics, a term coined by Ken Evans, are composites that become thicker perpendicular to the applied force when stretched, what leads to negative Poisson ratios [1,2]. Mechanical metamaterials [3,4] are periodic structures, counterpart of electromagnetic metamaterials, consisting of materials with high contrast in mechanical properties, which have been studied at small scales ranging from micrometers to centimeters. Control of surface seismic waves has been experimentally demonstrated in soils structured at the metric scale [5] that can be viewed as analogues of phononic crystals with holes [6]. In order to achieve control of seismic waves in the sub-wavelength regime, it seems natural to look at large scale analogues of mechanical [3,4] and acoustic [7-16] metamaterials, which are periodic structures that can manipulate acoustic (e.g. pressure) and elastic (e.g. surface Lamb, Rayleigh, or bulk coupled shear and pressure) waves. The Veselago-Pendry flat lens via negative refraction is perhaps the most famous paradigm of electromagnetic [17-19], acoustic [20] and platonic [21] crystals. An interesting application that arises from the periodic distribution of boreholes or inclusions embedded in a soil is as aforementioned the seismic wave shielding for Rayleigh wave frequencies within stop bands [5] (frequency intervals where, under certain conditions, propagation of elastic waves, or some of their polarizations, is forbidden), but also the flat seismic lens [22] for Rayleigh waves. Interestingly, forest of trees can also serve as seismic metamaterials for Rayleigh waves [23]. Such experiments on control of Rayleigh waves can find some applications in protection of urban infrastructures against earthquakes in soft sedimentary soils [24,25].

However, one would like to design seismic metamaterials that have the ability to create band gaps not only for surface Rayleigh waves [5,22,23] but also for all other elastic wave polarizations : within certain frequency ranges known as complete stop bands, an incoming mechanical wave would be completely reflected by the structure, whether the seismic wave signal propagates near the air-soil interface (Rayleigh or Love waves) or within the soil (coupled shear and pressure mechanical waves). Notably, it is an interesting challenge to achieve the base isolation of a structure by applying the concept of metamaterials in civil engineering. In what follows, we would like to demonstrate that base isolation with auxetic-like metamaterials could be an effective way to improve the seismic response of a building due to an earthquake in contrast to traditional seismic design methods which aimed to increase the strength of the structural parts of the building.

The Poisson ratio $v$ is a positive parameter for many common isotropic elastic natural materials, with values ranging from 0 (e.g. cork) to 0.5 (e.g. rubber), which means that they expand laterally upon pushing on them axially. Man-made auxetics were introduced when Roderik Lakes studied three-dimensional foams with $v < 0$ [1] in the mid eighties. More recently, there has been a keen interest in rationally designing auxetic mechanical metamaterials, which can additionally be intentionally anisotropic. However, auxetics need not be necessarily engineered, e.g. living bone tissue is a natural anisotropic auxetic material [2]. All fabricated rationally designed auxetic metamaterials are based on a few basic motifs [3]. One of the simplest motifs is the bow-tie element, which we shall consider in our study. Upon pushing (resp. compressing) along the vertical $z$ direction, a usual (i.e. not auxetic) material contracts (resp. expands) along the horizontal $x$ and $y$ directions. For a negative $v$, a compression along the vertical axis $z$ leads to a contraction of the shown bow-tie motif in figure 1 along the horizontal $x$ and $y$ direction [4].
In figure 1, the sign and magnitude of $v$ can be controlled via the angle α [4]. A zero Poisson ratio $v$ is expected to occur near $α = π/2$. The Poisson ratio effectively changes if the modulus of the strain is not small compared to unity, i.e., if one leaves the linear mechanical regime and enters the nonlinear regime, in which the change in the angle $α$ is no longer negligible [3]. This basic bow-tie motif can be assembled into two-dimensional model systems and into three-dimensional (anisotropic) mechanical metamaterials [4].

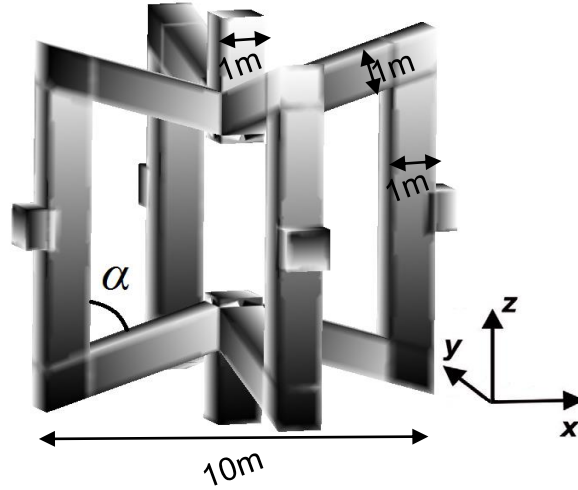

**Figure 1.** Illustration of the basic bow-tie element (known to lead to negative to Poisson's ratios in the static limit [4]), which is designed here to achieve tunable elastic stop bands for seismic waves (the periodic cell is 10 meter wide).

The angle $\alpha$ and hence the Poisson ratio $v$ can be adjusted in the fabrication process, as was demonstrated by the group of Martin Wegener at Karlsruhe Institute for Technology [4]. The quoted Poisson ratios $v$ measured by this group (more precisely, the components $v_{ij}$ of the Poisson matrix since the composite medium has an anisotropic elastic response) can be tailored to positive, near zero or even negative values. In the present paper, we propose to use such auxetic-like composites in order to design elastic stop band metamaterials for seismic waves propagating in sedimentary soils at hertz frequencies [5].

## 2. Stop band properties of auxetic-like metamaterials

We select concrete structures that should be fairly simple to fabricate and with elastic and geometric parameters that permit seismic waves molding. For the calculation of the band structure, the finite element method was used (Comsol Multiphysics). In the computations of figure 2, only one basic cell was considered with quasi-periodic Boundary Conditions enforced along each direction, in accordance with the Floquet-Bloch theorem. One usually starts with a vanishing Bloch vector **k** =(0,0,0), so that the resulting frequency spectrum consists of a well-defined countable set of eigenfrequencies associated with periodic eigenfields (stationary waves) sitting on local extrema of the band structure (sometimes at band gap edges). Changing the **k** vector value used in the quasi-periodic Boundary Conditions, one finds the eigenfrequencies and associated eigenfields that describe the overall band structure. In our calculations, we considered the irreducible Brillouin zone ΓXMR, where Γ=(0,0,0), X=($2\pi/a$,0,0), M=($2\pi/a,2\pi/a$,0) and R=($2\pi/a,2\pi/a,2\pi/a$), with *a* the array pitch. We realize that one should in principle describe all points within the irreducible zone (which is the volume of a tetrahedron), but it is customary to only look at its edges (invoking some symmetry reasons) for the sake of saving computational time.

The dimensions used for the computations of band structures associated with the bow-tie motifs shown in figure 2a)-d) are cells 10x10x10 m³ (same length along the *x, y* and *z* directions). Each cell is connected to its neighboring cells within the lattice via 6 bars are 1x1 m² in cross-section. The cross-section of each bar constituting the bow-tie element in the basic cell is also 1x1 m². The four cases studied in this work consist of bars made from concrete and the only thing that changes between the cases is the angle: $α = 0°$; $α = 30°$; $α = 45°$; $α = 60°$; Note that we use the density $ρ = 2.3$ g/cm³, the longitudinal (compressional) $c_l = 3.475$ km/s and the transverse (shear) $c_s = 2.022$ km/s components of elastic wave velocity for concrete. This corresponds to a Young modulus E=30 GPa and a Poisson ratio ν=0.3. Assuming that the lattice constant *a = 10 m,* frequencies are expected to be in the order of a few Hertz.

The first structure that was numerically examined is for an angle $α = 0°$ and is shown in fig. 2 a). In this case one can observe on the dispersion diagram that there are no stop bands in the frequency range 0 to 60 Hz. Stop bands actually occur at frequencies less relevant for seismic wave protection, i.e. above 50 Hz [5].

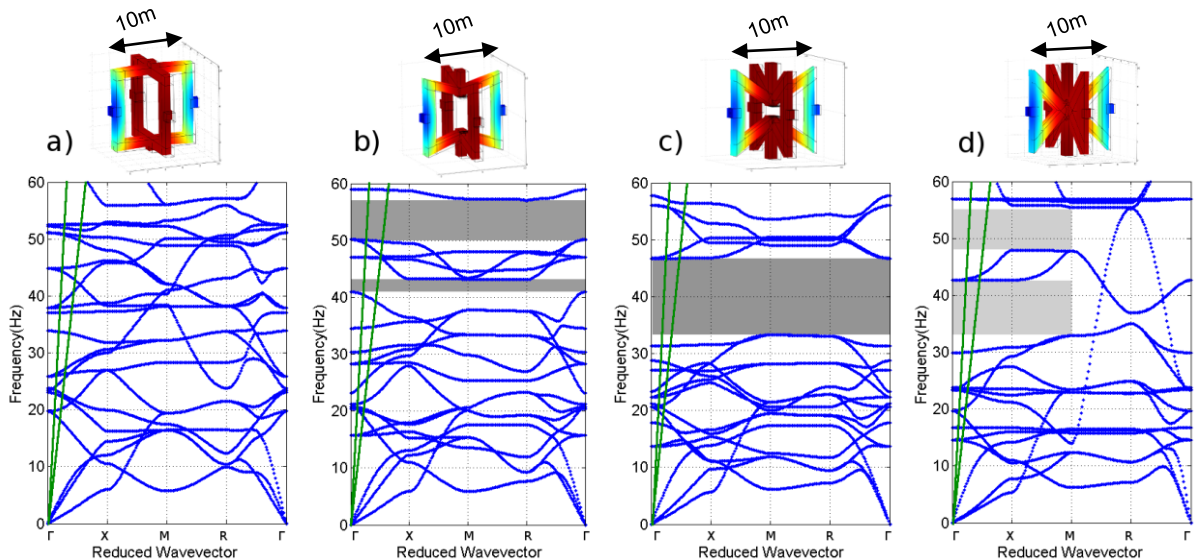

**Figure 2.** Band diagrams of auxetic-like metamaterials generated with elementary cells (10mx10mx10m) with bars (concrete) making an angle a) α = 0°; b) α = 30°; c) α = 45°; d) α = 60° where the horizontal axis is the normalized **k** wavevector describing the edges of the irreducible Brillouin zone ΓXMR and the vertical axis is the frequency (Hz). Green curves correspond to shear and pressure waves propagating in homogeneous isotropic bulk (concrete). Shaded regions mark the location of complete (dark grey) and partial (light grey) stop bands.

The second structure (for $α = 30°$, see fig. 2b)), which is in fact the first bow tie auxetic cell, gives rise to a tiny total band gap in a frequency range of interest for seismic wave protection, around 40 Hz. A second, larger, stop band lies above the frequency range of interest. This design could be used as an alternative to models of seismic protection proposed in [27,28].

The third structure (for $α = 45°$, see fig. 2c)), displays a wide total stop band that opens up around 40 Hz, from 33 Hz to 47 Hz (14 Hz of bandwidth, so with a relative bandwidth of 35%). By comparison, the fourth structure (for $α = 60°$, see fig. 2d) displays wide complete in-plane stop bands, which shrink along the MR and RΓ directions: This is easily understood by the fact that the vertical bars now meet at the center of the cell, making homogeneous vertical bars along which waves can travel at any frequency.

It then transpires from panels a)-d) that the auxetic-like structure with an angle of $\alpha = 45^o$ has the most promising band structure. Also worth noticing is the markedly different anisotropy of auxetic-like metamaterials in panels a)-d), wherein the second (resp. third) acoustic band displays decreasing (resp. increasing) group velocity with increasing angle $\alpha$. This might find some applications in seismic wave control such as cloaking.

### 3. Interpretation of stop bands with quasi-static and dynamic effective properties

There are interesting connections between models of metamaterials using low-frequency high-contrast and high-frequency low-contrast homogenization theories (see for instance chapter 1 in [29] where simple scalar wave equations are studied), in that the two routes lead to frequency dependent material parameters. Low-frequency high-contrast homogenization of the Navier equation is known to lead to a frequency dependent rank-2 tensor of effective density [30], but in what follows we would like to give some elements of proof for similar effective behaviour of Young's modulus and Poisson's ratio in the high-frequency regime. Let us start with some well-known facts of homogenization theory in the low-frequency low-contrast case. According to Christensen and Lo [31], and Nemat-Nasser and Willis [32] the effective density $<\rho>$ and Poisson's ratio squared $<\upsilon^2>$ can be approximated as follows in the case of a cubic array of ellipsoidal inclusions

$$<\upsilon^2> = \frac{\omega^2 a^2 <\rho>}{<C_{1111}>}, <\rho> = \rho^{(1)} f_1 + \rho^{(2)} f_2, \quad (1)$$

$$<C_{1111}> = \overline{C}_{1111}^{(1)} f_1 + \overline{C}_{1111}^{(2)} f_2, f_1 = 1 - f_2, f_2 = \frac{\pi b_1 b_2 b_3}{6a^3}, \quad (2)$$

where $\rho^{(1)}$ and $\rho^{(2)}$ are the densities of media 1 (bulk medium) and 2 (ellipsoidal inclusion), $\overline{C}_{1111}^{(1)}$ and $\overline{C}_{1111}^{(2)}$ are the first component of the elasticity tensor of media 1 and 2, respectively. Moreover, $f_1, f_2$ denotes the volume fraction of media 1 and 2, $a$ is the side length of the periodic cell, and $b_1, b_2, b_3$ are the semi-axes of the ellipsoidal inclusion. If we use these formulas to get a first estimate of the effective property of the structured medium of figure 1 in the quasi-static regime (leading order approximation), assuming that medium 1 is the volume occupied by the bars and medium 2 is the outer (air) medium, we cannot find any unusual behaviour, as testified by the lowest branches in the dispersion diagrams in figure 2a)-d). The fact that the pressure (third lowest) band on one hand and the shear (first and second lowest) bands on the other hand get increasingly different wavespeeds is due to the fact that in the quasi-static limit the auxetic-type metamaterial behaves effectively like a fluid.

However, the long wavelength limit breaks down at higher frequencies (or when the contrast gets bigger), where one needs to invoke dynamic effective properties that take into account the fine structure (here, bow-tie geometry) of the structured media, as described in [32-36]. Notably, in [35,36] it is clearly established that both the effective elasticity tensor $C_{eff}$ and the effective density $\rho_{eff}$ are frequency dependent, and [36] exhibits simultaneously negative values of entries of $C_{eff}$ and $\rho_{eff}$ in stop bands. This suggests negative effective Poisson's ratios might occur in many periodic structures in the stop bands, not necessarily with auxetic-type media such as first introduced by Lake [1], but also in many phononic crystals and mechanical metamaterials.

Indeed, if we assume that we are near resonances, then homogenization requires dynamic effective parameters that were first identified by Auriault in 1983 in the context of diffusion processes in high-contrast periodic media [37] using multiple scale expansion techniques, with subsequent extension to elastodynamics [30,38]. At the turn of the millennium, the mathematician Zikhov proposed a rigorous mathematical framework using two-scale convergence techniques in weak forms of scalar partial differential equations with fast oscillating parameters [39], that can be extended to elastodynamic equations [40]. Similar models of acoustic metamaterials exist using asymptotic analysis of Bessel functions [8], multi-structures [9], periodic unfolding techniques [10]. The link between the Poisson ratio and the appearance of stop bands in auxetic metamaterials, was pointed out in [41].

In our case, we consider a one-phase medium with stress free boundary conditions like in [41] but we carry out a more in-depth analysis by noticing that near resonances, modes get localised inside the bars of the structured medium in a way similar to what they would within a two-phase high-contrast medium $(\overline{C}_{1111}^{(2)} / \overline{C}_{1111}^{(1)} = O(\eta^2))$, where the stiff phase $\overline{C}_{1111}^{(1)}$ occupies the volume of the bars and the soft phase $\overline{C}_{1111}^{(2)}$ is the vacuum between the bars. This configuration of a composite with connected high-rigidity solid and soft inclusions has been studied by Auriault and Boutin in [30], using the usual ansatz in the displacement field $u = u^0(x,y) + \eta u^1(x,y) + \eta^2 u^2(x,y) + ...$ where $u^i(x,y) = u^i(x_1, x_2, x_3, y_1, y_2, y_3)$ are periodic fields in the microscopic variable y. They found that the effective medium is described by the following homogenized Navier equation:

$$div_x(\sigma_{eff}^x(u_{eff})) + \omega^2 \rho_{eff}^x(\omega) u_{eff} = 0, \qquad (3)$$

where $u_{eff}(x) = \int u^0(x,y) dy := \iiint (u_1^0, u_2^0, u_3^0)(x_1, x_2, x_3, y_1, y_2, y_3) dy_1 dy_2 dy_3$ is the homogenized displacement field, $\sigma_{eff}^x$ and $\rho_{eff}^x$ are the rank-2 effective stress and density tensors, respectively (both of which only depend upon the macroscopic variable x).

The effective (anisotropic) density is given by the following Drude-like expression

$$\rho_{eff}^x(\omega) = <\rho> I + \rho \sum_{i=1}^{\infty} \frac{<\phi^i> \otimes <\phi^i>}{<\int \phi^i \phi^i dy>} \frac{1}{\omega_i^2/\omega^2 - 1}, \qquad (4)$$

with $\phi^i(y) = (\phi_1^i, \phi_2^i, \phi_3^i)(y_1, y_2, y_3)$ a vector valued eigenfunction of the eigenvalue problem

$$div_y(C^y : \varepsilon^y(\phi^i)) = -\lambda_i \phi^i \text{ in the soft phase and } \phi^i = 0 \text{ on its boundary,} \qquad (5)$$

which is set on the microscopic periodic cell, which from renormalization is the unit cube $[0,1]^3$. It is well known that for this spectral problem there is a countable set of eigenvalues $0 \leq \lambda_1 \leq \lambda_2 \leq \lambda_3 \leq ... < +\infty$ with associated eigenfunctions $\phi_1^i, i = 1,2,3,...$ generating an orthogonal basis in the space of finite energy functions. When the frequency squared $\omega^2$ gets close to $\omega_i^2 = \lambda_i / \rho$ takes negative values, what can be used as an interpretation of stop bands as seen in figure 2 b)-d). A negative effective mass density could be notably invoked in stop bands of [42-45], but this would come short to a complete explanation of band structures.

Besides from the frequency dependent effective density in (3), the effective stress tensor is related to an effective elasticity tensor through $\sigma_{eff} = C_{eff} : \varepsilon^x$ where the tensor of deformation has the form $\varepsilon_{ij}^x = \frac{1}{E}\left[\sigma_{ij}(1+\upsilon) - \upsilon\delta_{ij}\sum_{k=1}^{3}\sigma_{kk}\right]$ with $\delta_{ij}$ the Kronecker symbol, E the Young modulus and ν the Poisson ratio. The effective elasticity tensor is computed from an annex problem of elasto-static type set on the unit cell excluding the soft phase by noticing that the stiff phase moves like a rigid body (translation) at the leading order of elastic field displacement. The effective elasticity tensor essentially contains the anisotropic features of the stiff phase, and it does not seem to have much to do with appearance of stop bands i.e. it is not frequency dependent. However, as recognised in [36], there is no uniqueness in the description of the homogenized medium at high frequencies, and alternative effective parameters can be achieved by assuming other types of limits in phase contrasts when one leaves the quasi-static regime. This is exactly what has been found by Richard Craster and coauthors in 2010 [33] when they introduced the concept of high-frequency homogenization (HFH). In their seminal paper, these authors take the usual ansatz for the displacement field, but instead of considering a high-contrast, they rescale the frequency and they are led to frequency dependent effective parameters for both stress and density tensors in the homogenized Navier equations [35]. Regarding HFH of thin plates, Tryfon Antonakakis and Craster find frequency dependent effective rigidity and density in the Kirchoff-Love equations [34], which is consistent with the work of Torrent et al. that demonstrates negative effective Young modulus, density and Poisson ratio using scattering matrix asymptotics in thin periodic plates with soft inclusions [46]. We thus claim that stop bands in figure 2 b), c), d) can be interpreted in terms of negative effective density, Young's modulus and Poisson ratio, depending upon the homogenization approach used. Classical work on auxetic materials [47-52,60] and mechanical metamaterials [53,61] as well as on homogenization [54-57] do not seem to have looked into the dynamic effective properties of auxetic-type materials, so we hope our section will foster theoretical and experimental studies in this direction.

4. **Quantification of energy loss through auxetic-like metamaterials**

As we discussed in the previous section, there are further dynamic effects induced by the effective elasticity tensor $C_{eff}$ that lead to other negative effective parameters, including the bulk modulus and Poisson ratio, as shown by Torrent, Pennec, Djafari-Rouhani in the context of a thin plate theory [46] that seems particularly well suited to the configuration of an auxetic-like metamaterial plate shown in figure 3. The detailed asymptotic analysis required for the identification of the frequency dependent effective elasticity tensor goes beyond the scope of the present paper but, one can clearly see the dramatic changes in the higher part of the dispersion diagrams in figure 2 a)-d) when the angle between bars changes, which is caused by a dynamic effective tensor changing radically of nature. Besides from that, there are marked local resonances in figure 3a) that can be interpreted with a negative effective Poisson ratio according to [46]. One should note in passing that the fact that there is a dynamic negative effective density does not necessarily mean that there is a negative effective Poisson ratio in 3D periodic structures, but we find that for plates the dynamic effective density, rigidity and Poisson ratio have the form of (4), which is consistent with [46], provided that the wave wavelength is much larger than the plate thickness. Furthermore, one should note that auxetic-like metamaterials exhibit slow modes in figure 2 (dispersion of pressure and shear waves propagating in the homogeneous concrete bulk are marked by green lines), and this can be interpreted in terms of vanishing effective density, Young's modulus and Poisson's ratio in light of HFH.

Let us now focus our attention on slow (localized) modes responsible for devastating buildings' resonances. In order to carry out their band structure analysis, we consider a model for a doubly periodic array of bars 80m high, with a square cross-section of 20m x 20m (that stand for buildings) lying atop a homogeneous concrete plate (40mx40mx40m). We enforce Floquet-Bloch conditions on the vertical sides of the plate, with stress free conditions elsewhere. The elastic band structure is shown in figure 3a). Note that the periodicity of the cell is irrelevant in the frequency range of interest since the band folding occurs far above 60 Hz (about 200 Hz for compressional waves). Eigenfrequencies of the buildings appear as flat bands on the dispersion diagram. One can also see that the first band on figure 3a) has a quadratic behaviour around the $\Gamma$ point, unlike in figure 2, which is due to the fact that the buildings lye atop a 40m deep concrete plate, so the corresponding elastic wave is a Lamb (flexural type) surface wave.

We display in figure 3b) a representative eigenmode at 17 Hz of a building which corresponds to the fifth flat band in figure 3a). This vibration is completely suppressed when the plate is structured with an auxetic-like metamaterial in figure 3c). The structured plate consists of a supercell made of $4^3$ elementary cells with bars making an angle $\alpha = 45°$ like in figure 2c) with Floquet Bloch conditions on vertical sides and stress-free conditions elsewhere. Upon inspection of the band structure corresponding to the configuration in figure 3c)-- which is not shown as it has a large number of bands-- we notice a stop band that opens around 17 Hz: This tiny stop band that is marked by two closely located local extrema in the transmission (black) curve near 20 Hz in figure 4a, can be interpreted in terms of a negative effective density, rigidity and Poisson's ratio: indeed, the building's vibration creates a monopolar resonance (a defect mode) within the auxetic supercell array, so that the Drude-like effective formula (4) applies to all of these parameters, see [46] for similar effective properties of a periodic plate.

We observe in figure 4, some low transmission from 0 to 15 Hz, which can be attributed to the liquid-like effective properties of the plate in the quasi-static limit (see section 3). Indeed, figure 2c) displays anisotropic bands with small velocities compared to a homogeneous isotropic plate, and the same holds true from the band structure associated with the supercell consisting of $4^3$ cells like in figure 2c). Physically speaking, by using auxetic-like elements we soften the medium underneath the building, which leads to a strong impedance mismatch between the auxetic-like building's foundation and the homogeneous plate medium. This results in a total reflection over a broad range of frequencies, according to the transmission loss of figure 4, that completely suppress the vibrations of the building at 1Hz and 9 Hz (marked by the first two local resonances in figure 3a). The suppression of the building's resonance at 37 Hz is more conventional, since this falls within a Bragg stop band according to figure 2c).

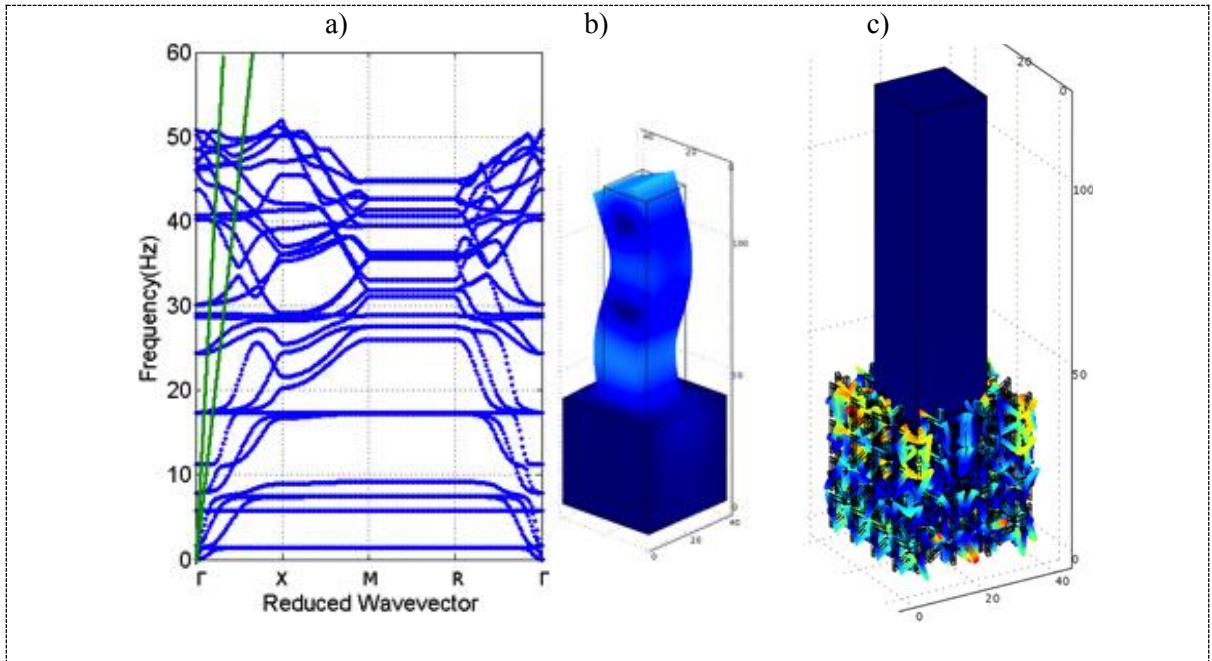

**Figure 3.** a) Band diagram for a concrete building (20mx20mx80m) atop a concrete homogeneous plate (40mx40mx40m); Floquet-Bloch conditions are set on vertical sides of the plate and stress-free conditions hold elsewhere; Flat bands correspond to eigenmodes of the building around 1 Hz, 9 Hz, 17 Hz etc. that couple to the flexural band (lowest band at $\Gamma$). b) Representative eigenmode of a) at 17 Hz. c) Eigenmode with suppressed building's vibration around 17 Hz when the plate is structured with auxetic metamaterial ($4^3$ elementary cells with bars making an angle as in Fig. 2 c)).

Let us now quantify the amount of elastic energy stored within the building and the transmission loss through this building, with and without the auxetic-like metamaterial. To do so, we apply periodic boundary conditions on the transverse sides of an elongated homogeneous concrete plate comprising a building (see figure b1)-d2)) so as to prevent appearance of transverse beam modes. A line source is applied at the left edge of the plate in order to simulate an incoming earthquake. The red (resp. black) curve in figure 4a) shows the transmission through the plate with a building on its own (resp. with the auxetic-like metamaterial). One can see that regardless of the frequency of the incoming wave, transmission through the building lying atop the auxetic-like metamaterial foundation is at least 10 dB lower than without the metamaterial. One should note that any building located behind the auxetic-like metamaterial would be protected. In order to quantify the level of protection of the building lying atop the auxetic-like metamaterial, we compute (magenta curve) the ratio of the norm of the total displacement field stored inside the building above the plate by the norm of the total displacement field stored inside the building above the plate structured with auxetic-like metamaterial. The building's protection which is mainly achieved via wave velocity impedance mismatch is clearly demonstrated throughout the frequency range 0 to 100 Hz, with a slightly more pronounced protection in the range of stop band frequencies from 33 to 47 Hz.

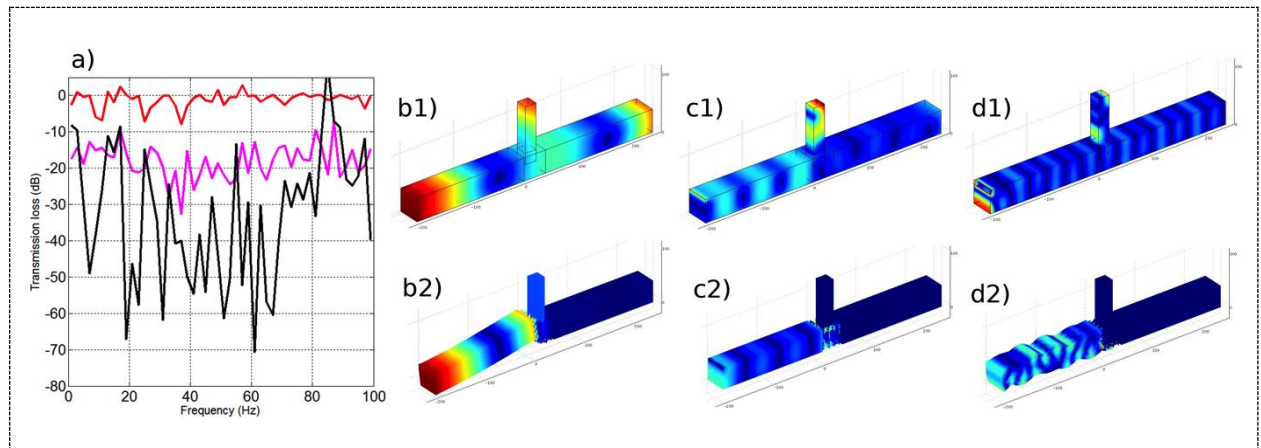

**Figure 4.** Quantification of energy loss a) with red (resp. black) curves representing transmission through a building atop a concrete plate (resp. a concrete plate structured with an auxetic metamaterial); Magenta curve gives the total displacement field stored inside the building above an auxetic-like metamaterial normalized with respect to the same total displacement field above the homogeneous plate. Representative out-of-plane displacement without (b1),c1) and d1) and with (b2),c2)) and d2) auxetic-like metamaterial illustrate how seismic protection works at 1Hz (b1),b2)) with the impedance mismatch between the homogeneous plate and the auxetic-like metamaterial ; at 9Hz (c1),c2)) with a local resonance ; at 37 Hz (d1),d2)) with the Bragg gap.

Some typical examples of out-of-plane (vertical) displacement field are shown at 1 Hz without b1) (resp. with b2)) auxetic-like metamaterial, at 9 Hz without c1) (resp. with c2)) auxetic-like metamaterial and at 37 Hz without d1) (resp. with d2)) auxetic-like metamaterial. The periodic foundation is clearly different from the traditional base isolation in which it causes a fundamental frequency shift in the structure, thus reducing its response and generating a frequency gap. Implementing auxetic-like cells in the foundations of sensitive buildings would significantly affect their static and dynamic responses during earthquakes.

## 5. Concluding remarks and perspectives on seismic metamaterials

The main conclusion of our study is that auxetic-like materials who have been widely studied for their special properties linked to a negative Poisson ratio ν, also have interesting stop band properties that can be used in the context of seismic wave protection. The stop bands can be associated with frequency dependent mass density ρ, but also bulk modulus [12,58], shear modulus, in a way similar to what was unveiled in acoustic metamaterials. Nonetheless, it is also possible to achieve frequency dependent Poisson's ratios in stop bands of certain mechanical metamaterials [36,46], what is less well known. Theories of low-frequency high-contrast homogenization [37-40] and high-frequency homogenization [33-35] exist that allow to investigate such dynamic effective properties of auxetic metamaterials. Using arrays of auxetic cells for seismic structural protection is a new way to perform elastic band gaps in order to prevent seismic wave propagation over specific frequency ranges, but we also discovered that the very strong impedance mismatch between elastic wave velocity within the homogeneous bulk of concrete and that within the auxetic metamaterial enables suppression of wave transmission over a very large frequency range (0 to 100 Hz), which opens interesting avenues in earthquake protection. We stress that we have performed our study in elastic conditions, but we are well aware that soils display visco-elastic properties causing frequency dependent damping effects in soft sediments [59].

One could also envisage to combine auxetic-like seismic metamaterials within the soil with other types of protections such as forests of trees [23] to further widen the range of stop band frequencies for Rayleigh waves. It might be interesting to also investigate thick pillars above the soil, since previous numerical and experimental studies on small scale photonic crystals composed of nickel pillars grown on a lithium niobate substrate [14-15] have shown complete stop bands at low frequencies which are also robust versus disorder. Such seismic metamaterials would offer alternative ways to protect buildings, which necessitate specific modal analysis [24,59] by civil engineers in order to avoid disastrous seismic site effects.